\documentclass{mem}
\usepackage{natbib}\usepackage{txfonts}\usepackage{balance}
\usepackage{graphicx}
\usepackage[a4paper,breaklinks,dvipdfm]{hyperref}
\idline{75}{282}
\begin{document}
\def\teff{$T\rm_{eff }$}
\def\kms{$\mathrm {km s}^{-1}$}

\title{
Dynamics of tidally captured planets in the Galactic Center
}

   \subtitle{}

\author{
A.A. \,Trani\inst{1,2}
\and M. \, Mapelli\inst{2}
\and M. \, Spera\inst{2}
\and A. \, Bressan\inst{1,2}
	      }

\newcommand{\msun}{\,\mathrm{M}_\odot} 
\newcommand{\mjup}{\,\mathrm{M}_{\rm Jup}}
\newcommand{\yr}{\,\mathrm{yr}} 
\newcommand{\pc}{\,\mathrm{pc}}
\newcommand{\AU}{\,\mathrm{AU}} 

\institute{
Scuola Internazionale Superiore di Studi Avanzati, via Bonomea 265,
I-34136, Trieste, Italy,
\email{aatrani@gmail.com}
\and
Istituto Nazionale di Astrofisica --
Osservatorio Astronomico di Padova, Vicolo Osservatorio 5, 
I-35122 Padova, Italy
}

\authorrunning{Trani et al. }

\titlerunning{Planetary dynamics in the GC}

\abstract{
Recent observations suggest ongoing planet formation in the innermost parsec of our Galaxy. The super-massive black hole (SMBH) might strip planets or planetary embryos from their parent star, bringing them close enough to be tidally disrupted. We investigate the chance of planet tidal captures by running three-body encounters of SMBH-star-planet systems with a high-accuracy regularized code. We show that tidally captured planets have orbits close to those of their parent star. We conclude that the final periapsis distance of the captured planet from the SMBH will be much larger than $\sim{}200$ AU, unless its parent star was already on a highly eccentric orbit.
}
\maketitle{}

\section{Introduction}
Recent radio continuum observations suggest the presence of photoevaporating protoplanetary disks in the innermost $\sim{}0.1$ pc of the Galactic Center (GC, \citealt{yus15}), very close to the super-massive black hole (SMBH). 
\citet{map15} showed that rogue planets are too faint to be observed in the GC with current facilities, unless their near-infrared luminosity is enhanced by photoevaporation combined with tidal disruption induced by the SMBH.

A protoplanetary origin has been suggested for the dusty object G2, which has been observed orbiting the SMBH on an highly eccentric orbit ($e \sim 0.98$) with very small periapsis ($a \sim 200 \AU$, \citealt{gil12}): \citet{mur12} proposed that G2 is a low-mass star with a proto-planetary disk, while \citet{map15} suggested that the properties of G2 are consistent with a planetary embryo tidally captured by the SMBH.

In this proceeding, we study the possibility that planets and protoplanets are tidally captured by the SMBH.

\section{Methods}
We ran $10^4$ simulations of a three-body hierarchical system composed of a SMBH, a star and a planet initially bound to the star. We use a fully regularized N-body code that implements the Mikkola's algorithmic regularization \citep{mikkola1999a}. 
The orbit of the star around the SMBH is modeled following the properties of the stars in the clockwise (CW) disk, the stellar disk observed at $\sim 0.1 \pc$ from the SMBH \citep{do13,yelda2014}. 
We assume the initial orbit of the planet around the star is circular (with semi-major axis $\sim{}10-100$ AU), coplanar and prograde with respect to the star orbit. More details about our simulations will be provided in Trani et al., in preparation. 
%
\begin{figure}[t!]
	\centering
	\resizebox{1.\hsize}{!}{\includegraphics[clip=true]{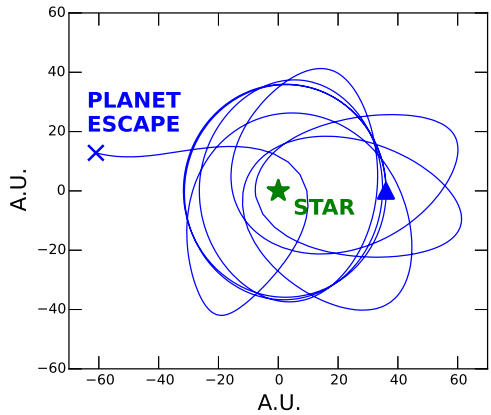}}\hspace{20pt}
	\caption{\footnotesize
		Planet trajectory in the reference frame that corotates with the star in one of our simulations. The SMBH position is always along the negative $x-$axis. Blue line: planet trajectory. Blue triangle: initial planet position. Blue cross: planet position at the time the planet becomes unbound with respect to the star. Green star: star position.} 
	\label{fig:gorb}
\end{figure}

\section{Results and conclusions}
Figure~\ref{fig:gorb} shows the trajectory of one of the simulated planets. The  planet completes several orbits around the star, even if its motion is strongly perturbed by the tidal field of the SMBH. As the star approaches its periapsis, the planet is tidally captured by the the SMBH. 

\begin{figure}[t!]
	\centering
	\resizebox{1.\hsize}{!}{\includegraphics[clip=true]{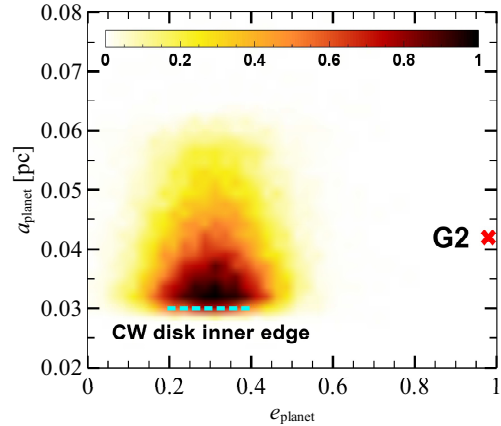}} 
	\caption{\footnotesize
		Cumulative probability map of semi-major axis and eccentricity of the simulated planets. Red cross: G2 cloud. Cyan dotted line: inner edge of the CW disk.}
	\label{fig:gorb2}
\end{figure}

Figure \ref{fig:gorb2} shows the cumulative probability map of finding an unbound planet in the semi-major axis -- eccentricity phase space. No planet can match the orbits of the G2 cloud. In particular, none of the simulated planets can achieve a highly-eccentric orbit.
In fact, the closest periapsis passage of an unbound planet in our simulations is $1750\rm\, AU$, a factor of $\sim 9$ larger than the periapsis passage of the G2 cloud.

We speculate that perturbations from other stars in the disk may bring planets into nearly-radial orbits. In forthcoming studies we will investigate the effect angular momentum transport and scatterings with other stars on the dynamics of planets in the disk.

\begin{acknowledgements}
AB, AT and MM acknowledge financial support from INAF through grant PRIN-2014-14. MM acknowledges financial support from MIUR through grant FIRB 2012 RBFR12PM1F and from the MERAC Foundation.
\end{acknowledgements}


\end{document}